\documentclass[12pt]{article}

\usepackage{times}
\usepackage{fullpage}
\usepackage{amsfonts,amssymb,amsmath}
\usepackage{latexsym}
\usepackage{url}
\usepackage[pdftex]{color}
\usepackage[breaklinks]{hyperref}

\newcommand{\R}{\mathbb{R}}

\newcommand{\braket}[2]{\langle #1, #2 \rangle}
\def\01{\{0,1\}}
\newcommand{\OR}{\mathrm{OR}}
\newcommand{\AND}{\mathrm{AND}}

\newcommand{\ADV}{\mathrm{ADV}}

\newcommand{\ignore}[1]{}

\newtheorem{theorem}{Theorem}
\newtheorem{lemma}[theorem]{Lemma}

\newtheorem{definition}[theorem]{Definition}

\newcommand{\thmref}[1]{\hyperref[#1]{{Theorem~\ref*{#1}}}}
\newcommand{\lemref}[1]{\hyperref[#1]{{Lemma~\ref*{#1}}}}
\newcommand{\corref}[1]{\hyperref[#1]{{Corollary~\ref*{#1}}}}
\newcommand{\eqnref}[1]{\hyperref[#1]{{Equation~(\ref*{#1})}}}
\newcommand{\claimref}[1]{\hyperref[#1]{{Claim~\ref*{#1}}}}
\newcommand{\remarkref}[1]{\hyperref[#1]{{Remark~\ref*{#1}}}}
\newcommand{\propref}[1]{\hyperref[#1]{{Proposition~\ref*{#1}}}}
\newcommand{\factref}[1]{\hyperref[#1]{{Fact~\ref*{#1}}}}
\newcommand{\defref}[1]{\hyperref[#1]{{Definition~\ref*{#1}}}}
\newcommand{\exampleref}[1]{\hyperref[#1]{{Example~\ref*{#1}}}}
\newcommand{\hypref}[1]{\hyperref[#1]{{Hypothesis~\ref*{#1}}}}
\newcommand{\secref}[1]{\hyperref[#1]{{Section~\ref*{#1}}}}
\newcommand{\chapref}[1]{\hyperref[#1]{{Chapter~\ref*{#1}}}}
\newcommand{\apref}[1]{\hyperref[#1]{{Appendix~\ref*{#1}}}}

\newenvironment{proof}[1][Proof: ]
{\noindent {\bf #1}}
{{\hfill $\Box$}\\
 \smallskip}

\begin{document}
\title{A note on the sign degree of formulas}
\author{Troy Lee \thanks{Rutgers University.  Work conducted while at Columbia University.}}
\date{}
\maketitle

\begin{abstract}
Recent breakthroughs in quantum query complexity have shown that any formula
of size $n$ can be evaluated with $O(\sqrt{n}\log(n)/\log \log(n))$ many
quantum queries in the bounded-error setting 
\cite{FGG08, ACRSZ07, RS08b, Rei09}.  In
particular, this gives an upper bound on the approximate polynomial degree of formulas
of the same magnitude, as approximate polynomial degree is a lower bound on quantum
query complexity \cite{BBCMW01}.

These results essentially answer in the affirmative a conjecture of
O'Donnell and Servedio \cite{O'DS03} that the sign degree---the minimal degree
of a polynomial that agrees in sign with a function on the Boolean cube---of
every formula of size $n$ is $O(\sqrt{n})$.

In this note, we show that sign degree is super-multiplicative under
function composition.  Combining this result with the above mentioned
upper bounds on the quantum query complexity of formulas allows the
removal of logarithmic factors to show that the sign degree of every
size $n$ formula is at most $\sqrt{n}$.
\end{abstract}

\section{Introduction}
There is a growing body of work which uses techniques of quantum
computing and information to prove results whose statements have
no reference to quantum at all \cite{KW04, Aar03, Aar05, LLS06, Wol08}.  One simple
application of this type is to the construction of low-degree polynomials that approximate a
Boolean function.  Beals et al. \cite{BBCMW01} show that one-half the minimum degree of a
polynomial which approximates a function $f$ on the Boolean cube within error $1/3$ (in terms of
$\ell_\infty$ norm) is a lower bound on the $1/3$-error quantum query
complexity of $f$.  Turning this around, if $f$ has a $d$-query bounded-error quantum
algorithm, then it has approximate polynomial degree at most $2d$.  Using
quantum algorithms has proven a remarkably powerful means of constructing
approximating polynomials, and in quite a few cases no other construction is
known, for example \cite{BNRW07}.

Another example where quantum algorithms show new bounds on approximate
degree is in the case of functions described by small formulas.  A formula is a binary
tree where internal nodes are labeled by
binary $\AND_2$ or $\OR_2$ gates and leaves are labeled either by a literal
$x_i$ or its negation $\neg x_i$.  The size of a formula is the number of
leaves.  Recent breakthroughs in quantum query complexity have shown that if
a function $f$ can be computed by a formula of size $n$, then there is a
quantum query algorithm that can evaluate $f$ with high probability in
$O(\sqrt{n} \log(n)/ \log \log(n))$ many queries
\cite{FGG08, ACRSZ07, RS08b, Rei09}.  By the above connection, this implies
that the approximate polynomial degree of any formula is also
$O(\sqrt{n} \log(n)/ \log \log(n))$.  Previous to these results,
it was an open question, raised by O'Donnell and Servedio \cite{O'DS03}, to
show that every size $n$ formula has {\em sign} degree $O(\sqrt{n})$.  The
sign degree of $f$, denoted $\deg_\infty(f)$, is the minimum degree of a
polynomial which agrees in sign with $f$ for all $x \in \{-1,+1\}^n$.

In this note, we show a lemma about sign degree under function
composition.  Namely, if $f \circ g^n(x)=f(g(x^1), \ldots, g(x^n))$,
where $x=(x^1, \ldots, x^n)$, then
\[
\deg_\infty(f \circ g^n) \ge \deg_\infty(f) \deg_\infty(g).
\]
This lemma is often not tight: for example both $\AND_n$ and $\OR_n$
have sign degree one, whereas Minsky and Papert show that $\OR_n
\circ \AND_{n^2}^n$ has degree $n$.  When combined with the results
of Reichardt \cite{Rei09}, however, this lemma allows the removal of
log factors to fully resolve the question of O'Donnell and Servedio
and show that every size $n$ formula has sign degree at most
$\sqrt{n}$.  This upper bound is exactly tight for infinitely many
values of $n$ since for any $n=2^{2k}$, the parity function over
$\sqrt{n}$ variables is computed by a size $n$ formula and has sign
degree exactly $\sqrt{n}$.

\section{Preliminaries}
Let $[n]=\{1,2, \ldots, n\}$.  For a set $T \subseteq [n]$, we associate
the character $\chi_T: \{-1,+1\}^n \rightarrow \{-1,+1\}$ where
$\chi_T(x)=\prod_{i \in T} x_i$ for $x \in \{-1,+1\}^n$.  Every function
$f: \{-1,+1\}^n \rightarrow \{-1,+1\}$ has a unique expansion as a multilinear
polynomial
\[
f(x)=\sum_{T \subseteq [n]} \hat f_T \chi_T(x).
\]

The {\em polynomial degree} of $f$, denoted $\deg(f)$,  is the size of a
largest set $T$ for which $\hat f_T \ne 0$.  We say that $f$ has
{\em pure high degree} $d$ if $\hat f_T =0$ for all sets $T$ with $|T| < d$.

Our main object of study is the degree of polynomials which approximate a
function $f$.
\begin{definition}
Let $f: \{-1,+1\}^n \rightarrow \{-1,+1\}$.  For $\alpha \ge 1$ the
$\alpha$-{\em approximate degree} of $f$ is
\[
\deg_\alpha(f)=\min_p \left\{\deg(p): 1\le p(x)f(x) \le \alpha \
\mbox{ for all } x \in \{-1,+1\}^n\right\}.
\]
{\em Sign degree} is defined as
$$
\deg_\infty(f)=\min_p \left\{\deg(p): 1\le p(x)f(x) \
\mbox{ for all } x \in \{-1,+1\}^n\right\}.
$$
\end{definition}

Notice that for a fixed degree $d$ and approximation parameter $\alpha$
(possibly $\alpha=\infty$), determining if $\deg_\alpha(f)$ is at most $d$ can
be checked by determining the feasibility of a linear program.  On the other
hand, showing that the dual of this linear program is feasible implies
that $\deg_\alpha(f) > d$.  We encapsulate the
feasibility conditions of this dual program in the next lemma.

\begin{lemma}
Fix $1 \le \alpha \le \infty$ and let $f:\{-1,+1\}^n \rightarrow \{-1,+1\}$.
There exists a function $p: \{-1,+1\}^n \rightarrow \R$ such that
\begin{enumerate}
  \item $\braket{f}{p} \ge \begin{cases}
     \frac{\alpha -1}{\alpha+1} & \mbox{ if } \alpha < \infty \\
     1 & \mbox{ if } \alpha=\infty
     \end{cases}.$
  \item $\ell_1(p)=1$.
  \item $\braket{p}{\chi_T}=0$ for any character $\chi_T$ with
$|T| < \deg_\alpha(f)$.
\end{enumerate}
\label{dual_poly}
We refer to $p$ as a dual witness for $\deg_\alpha(f)$.
\end{lemma}

\section{Composition lemma} \label{sec:composition}
Let $f$ be a function $f: \{-1,+1\}^n \rightarrow \{-1,+1\}$, and
$g:\{-1,+1\}^m \rightarrow \{-1,+1\}$.  We define the composition of $f$ and
$g$ as $f \circ g^n:\{-1,+1\}^{mn} \rightarrow \{-1,+1\}$ where
$(f\circ g^n)(x)=f(g(x^1), \ldots, g(x^n))$ for $x=(x^1, \ldots, x^n)$.

Our composition lemma states that
$\deg_\infty(f \circ g^n) \ge \deg_\infty(f) \deg_\infty(g)$.
This lemma is often not tight---for example, both $\OR_n$ and $\AND_n$ have
sign degree $1$.  On the other hand, Minsky and Papert show that
$\OR_n \circ \AND_{n^2}^n$ has sign degree $n$.
Extending such a composition lemma to the bounded-error case, where it would
be nearly tight, would be a major breakthrough.  In particular, such a result
would resolve the approximate polynomial degree of the function
on $n^2$ many variables $\OR_n \circ \AND_n^n$, which is currently only known to be
somewhere between $n^{2/3}$ and $n$ \cite{AS04, HMW03}.

\begin{lemma}
Let $f$ be a function $f: \{-1,+1\}^n \rightarrow \{-1,+1\}$, and
$g:\{-1,+1\}^m \rightarrow \{-1,+1\}$.  Then for $1 \le \alpha \le \infty$
$$
\deg_\alpha(f \circ g^n) \ge \deg_\alpha(f) \deg_\infty(g).
$$
\label{lemma:main}
\end{lemma}

\begin{proof}
Let $p,q$ satisfy the conditions of \lemref{dual_poly} for
$\deg_\alpha(f)$ and $\deg_\infty(g)$, respectively.  If
$\deg_\infty(g)=0$ then the statement is trivial, so we assume that
$\deg_\infty(g) \ge 1$ and so $\braket{\chi_\emptyset}{q}=0$.
Notice that as $\ell_1(q)=\braket{g}{q}=1$ we must have $g(x)q(x)
\ge 0$ for all $x \in \{-1,+1\}^m$.  Thus we may express $q$ as
$q(x)=g(x) \mu(x)$ where $\mu(x) \ge 0$ for all $x$.

Define
\[
h(x)=2^n p(g(x^1), \ldots, g(x^n)) \cdot \prod_i \mu(x^i).
\]
Let us verify that $h$ has the properties of a dual witness.
\begin{align*}
\braket{f \circ g^n}{h}&= 2^n \sum_x f(g(x^1),\dots,g(x^n))
p(g(x^1),\dots,g(x^n)) \cdot \prod_i \mu(x^i)\\
&= 2^n \sum_{z \in \{-1,+1\}^n} f(z)p(z) \sum_{\substack{x: \\
g(x^i)=z_i \ \forall i}} \prod_i \mu(x^i)\\
 &=2^n \sum_{z \in \{-1,+1\}^n} f(z)p(z) \prod_{i}
\sum_{\substack{x^i: \\ g(x^i)=z_i}} \mu(x^i) \\
&= \sum_{z \in \{-1,+1\}^n} f(z) p(z) \\
&\ge \begin{cases}
\frac{\alpha-1}{\alpha+1} & \mbox{ if } \alpha < \infty \\
1 & \mbox{ if } \alpha=\infty
\end{cases}.
\end{align*}
The fourth equality holds since $\braket{\chi_0}{q}=0$ and
$\ell_1(q)=1$ imply
\[
\sum_{y: g(y)=-1} \mu(y)= \sum_{y:g(y)=1} \mu(y)=\frac{1}{2}.
\]
Next we verify that $\ell_1(h)=1$.  This follows quite similarly:
\begin{align*}
\ell_1(h)&= 2^n \sum_{z \in \{-1,+1\}^n} |p(z)| \prod_{i}
\sum_{\substack{x^i \\ g(x^i)=z_i}} |\mu(x^i)| \\
&=\sum_{z \in \{-1,+1\}^n} |p(z)|=1.
\end{align*}
Finally, we check that $h$ is orthogonal to all characters of degree less than
$\deg_\alpha(f) \deg_\infty(g)$.  To see this, write out
\begin{align*}
{\frac {h(x)}{2^n}}&=p(g(x^1), \ldots, g(x^n)) \cdot \prod_{i=1}^n \mu(x^i) \\
&= \sum_T \hat p_T \prod_{i \in T} g(x^i) \cdot \prod_{i=1}^n \mu(x^i) \\
&= \sum_{T:|T| \ge \deg_\alpha(f)} \hat p_T \left(\prod_{i \in T} g(x^i)
\mu(x^i) \cdot \prod_{j \not \in T} \mu(x^j) \right) \\
&= \sum_{T: |T| \ge \deg_\alpha(f)} \hat p_T \left(\prod_{i \in T} q(x^i)
\cdot \prod_{j \not \in T} \mu(x^j)
\right).
\end{align*}
For each fixed $T$, the term $\prod_{i \in T} q(x^i)$ is a product of at least
$\deg_\alpha(f)$ many polynomials $q(x^i)$ which are over disjoint sets of
variables, and each of which has pure high degree $\deg_\infty(g)$.  Thus the
product has pure high degree at least $\deg_\alpha(f) \cdot \deg_\infty(g)$.
Multiplying by $\prod_{j \notin T} \mu(x^j)$, which is a polynomial over
another set of disjoint variables, cannot decrease the pure high degree.
So the pure high degree of $h$ is at least
$\deg_\alpha(f) \cdot \deg_\infty(g)$.
\end{proof}

\section{Sign degree of formulas}
We now see how \lemref{lemma:main} can be used in conjunction with recent
results of Reichardt \cite{Rei09} to show that every formula of size $n$ has
sign degree at most $\sqrt{n}$.  The result of Reichardt we need shows that
the negative adversary bound characterizes quantum query complexity amortized over function composition.  The negative adversary bound \cite{HLS07} is a lower bound
technique for quantum query complexity which generalizes
the quantum adversary method of Ambainis \cite{Amb02, Amb03}, in particular the spectral
formulation of the adversary bound due to Barnum, Saks, and Szegedy \cite{BSS03}.

\begin{definition}[Negative adversary bound]
Let $f:\{-1,+1\}^n \rightarrow \{-1,+1\}$.  For each $i=1, \ldots, n$ let
$D_i$ be a zero-one valued matrix with rows and columns labeled by
$n$-bit strings and where $D_i[x,y]=1$ if $x_i \ne y_i$ and $D_i[x,y]=0$ otherwise.
Let $F$ be a zero-one valued matrix where $F[x,y]=1$ if $f(x)\ne f(y)$ and $F[x,y]=0$
otherwise.  Define
\[
\ADV^{\pm}(f)=\max_{\Gamma \ne 0} \frac{\|\Gamma \circ F\|}{\max_i \|\Gamma \circ D_i\|}.
\]
Here $\|A\|$ denotes the spectral norm of the matrix $A$.
\end{definition}

\begin{theorem}[Reichardt \cite{Rei09}]
For any function $f: \{-1,+1\}^n \rightarrow \{-1,+1\}$, let $f^{(k)}$ denote $f$
composed with itself $k$ times.  Then
$$
\lim_{k \rightarrow \infty} Q(f^{(k)})^{1/k}=\ADV^{\pm}(f).
$$
\end{theorem}

It is known that if $f$ has formula size $n$ then $\ADV^\pm(f) \le \sqrt{n}$.
This can be seen using the fact that
$\ADV^{\pm}(\AND_n)=\ADV^\pm(\OR_n)=\sqrt{n}$ and that
$\ADV^\pm(f \circ g) \le \ADV^\pm(f) \ADV^\pm(g)$ the adversary bound
is sub-multiplicative under function composition \cite{Rei09}.
Thus we have
$$
\deg_\infty(f)\le \lim_{k \rightarrow \infty} \deg_\infty(f^{(k)})^{1/k}
\le \lim_{k \rightarrow \infty} (2Q(f^{(k)}))^{1/k}\le \sqrt{n}.
$$
where the first inequality follows from the composition lemma of
Section~\ref{sec:composition} and the second is by the bound of
Beals et al. mentioned in the Introduction.

\section{Conclusion}
As with all classical results proven via quantum techniques, it would be
interesting to come up with a more direct proof.
For both the case of sign degree and quantum query complexity, the more
difficult case is formulas which are highly unbalanced.
For the complete AND-OR binary tree of size $n$,
one can quite easily give an explicit sign representing polynomial of degree
$\sqrt{n}$.  The benefit of the composition lemma seems to be that it reduces
the problem of showing an upper bound on the sign degree of $f$ to showing an
upper bound on the sign degree of $f^{(k)}$, which intuitively is a more
``balanced'' function.  While in the quantum case there is a good notion of
``approximately balanced'' to make this plan work (see \cite{ACRSZ07}), it still remains
to come up with a good classical notion of approximately balanced to push
such a proof through.

\section*{Acknowledgments}
I would like to thank Ben Reichardt for many discussions and for sharing a preliminary version 
of \cite{Rei09}, Rocco Servedio for conversations about this work and helpful 
comments on the writeup, and Ronald de Wolf for a careful reading of the writeup.


\newcommand{\etalchar}[1]{$^{#1}$}

\end{document}